\documentclass[aps,onecolumn,showpacs,groupedaddress]{revtex4-1}
\usepackage{amssymb}
\usepackage{mathrsfs}
\usepackage{pstricks}
\usepackage{color}
\usepackage{graphicx}
\usepackage[fleqn]{amsmath}
\usepackage{slashed}
\usepackage{amsmath}
\usepackage{mathtools}
\usepackage{array}
\usepackage{tabularx}
\usepackage{physics}
\usepackage{slashed}
\usepackage[utf8]{inputenc}
\usepackage[T1]{fontenc}
\usepackage{float}

\begin{document}


\title{\bf Nonideal Statistical Field Theory at NLO}


\author{P. R. S. Carvalho}
\email{prscarvalho@ufpi.edu.br}
\affiliation{\it Departamento de F\'\i sica, Universidade Federal do Piau\'\i, 64049-550, Teresina, PI, Brazil}





\begin{abstract}
In this work we introduce a field theory capable of describing the critical properties of nonideal systems undergoing continuous phase transitions beyond the leading order radiative corrections or in the number of loops (effective field theories limited only to leading order were recently defined in literature). These systems present defects, inhomogeneities and impurities as opposed to ideal ones which are perfect, homogeneous and pure. We compute the values of the critical exponents of the theory beyond leading order and compare with their corresponding experimental measured results. The results show some interplay between nonideal effects and ﬂuctuations. 
\end{abstract}


\maketitle


\section{Introduction}

\par Renormalization has played a central role for determining what theory we have to employ in describing consistently many-particle systems behavior at their deepest or fundamental level. In fact, today, such a set of fundamental theories for studying elementary particles and their interaction carriers, \emph{i. e.} the electromagnetic, strong and weak interactions composes what it is called the standard model of elementary particles and fields \cite{Peskin}. All these theories are renormalizable and the effects probed by them can be studied by taking into account fluctuations of the respective quantum fields at all scales of length or loop orders, thus turning them fundamental theories. For example, in quantum electrodynamics (QED) the anomalous magnetic moment of the electron, in the pure-QED sector, was computed up to four-loop order \cite{PhysRevD.73.013003}. But before achieving such renormalizable theories, the definition of earlier non-renormalizable or not fundamental ones (also called effective theories, which take fluctuations into account only up to leading loop order which is commonly the one-loop order) has helped to obtain the theories aforementioned. In Ref. \cite{Schwartz}, examples of non-renormalizable or effective theories as the Schrödinger equation, the Fermi theory of weak interactions, the chiral perturbation theory as the low-energy theory of pions and the general relativity as the low-energy theory of gravity are discussed, correspond to the four fundamental forces of nature, namely, the electromagnetic, weak, strong and gravitational interactions, respectively. The results obtained through these theories were limited only up to leading order (LO) in the number of loops, generally up to one-loop order \cite{Schwartz}. When the one-loop order term of some physical quantity vanishes, its leading order term is the two-loop one if this term is nonvanishing and so on. In fact, until the corresponding consistent renormalizable theories were found, the referred non-renormalizable theories were applied for practical calculations although not providing results valid for all scales of length or loop orders and were both not numerically precise and conceptually wrong. The results were expressed only as a one-loop order contribution while the real effects approached are a result of a sum of so many terms like one-, two-, three-loop order terms and so on. So it was necessary to find out some theories capable of furnishing results composed of one- and high loop orders and the non-renormalizable theories were discarded as fundamental theories and were employed at most as effective ones. 

\par In recent years, field theories designed to describe the critical properties of nonideal systems undergoing continuous phase transitions were defined \cite{CARVALHO2024139487,CARVALHO2023138187,ALVES2023138005,CARVALHO2023137683}. These systems present defects, inhomogeneities and impurities as opposed to ideal ones which are perfect, homogeneous and pure \cite{Wilson197475}. Such theories furnished results for the critical exponents describing the referred effects limited just up to leading loop order, through some parameter \cite{CARVALHO2024139487,CARVALHO2023138187,ALVES2023138005,CARVALHO2023137683}, and thus are characterized as effective theories. Taking the appropriate limit, all the results for ideal systems were recovered. Unfortunately, these theories are non-renormalizable and the results, although numerically incorrect or correct in some cases, once we can find a value of the parameter to match a given critical exponent value are, at least, conceptually wrong since they were written just as the sum of a nonideal LO and ideal higher order contributions (this hypothesis, namely of writing the critical exponents values just as the sum of a nonideal leading order and ideal higher ones, is not rigorous and since such theories are non-renormalizable the rigorous forms of writing the critical exponents is just up to nonideal LO and are shown in Appendix \ref{Critical exponents for some effective theories} for the corresponding theories \cite{CARVALHO2024139487,CARVALHO2023138187,ALVES2023138005,CARVALHO2023137683} and some other ones). On the other hand, the nonideal physical effects are a result of a sum of so many nonideal terms (not just up to the leading one), namely up to leading, next-to-leading, next-to-next-to leading order terms and so on. So the non-renormalizable or effective theories shown in Appendix \ref{Critical exponents for some effective theories} must be discarded as fundamental ones, \emph{i. e.} as ones representing defects, inhomogeneities and impurities taken at all scales of length and have to be employed at most as effective theories. To supply the lack of a fundamental theory that fulfills the task of taking nonideal effects beyond LO into account, we present some one.

\par Now we introduce renormalization group and $\epsilon$-expansion techniques capable of describing the referred effects at scales of length beyond LO. We present the values of the corresponding critical exponents beyond LO, namely, up to next-to-leading order (NLO) and compare with experimental measured results for systems with defects, inhomogeneities and impurities. As the present theory describes nonideal systems undergoing continuous phase transitions, we call such a theory as nonideal statistical field theory (NISFT) and it generalizes the previous theory for ideal systems \cite{Wilson197475}. As the theory presented in this work represents O($N$)$_{a}$ $\lambda\phi^{4}$ theories, where $a$ is the parameter taking into account to the nonideal properties, it emerges naturally new generalized O($N$)$_{a}$ universality classes each one characterized by the $a$ parameter. When $a\rightarrow 1$, we recover the ideal results \cite{Wilson197475}. Some specific models are: self-avoiding random walk ($N=0$), Ising ($N=1$), XY ($N=2$), Heisenberg ($N=3$), spherical ($N \rightarrow \infty$) models etc \cite{Pelissetto2002549}. Now we present the results for the critical exponents.

\begin{figure}[H]
\centering
\includegraphics[width=0.45\textwidth]{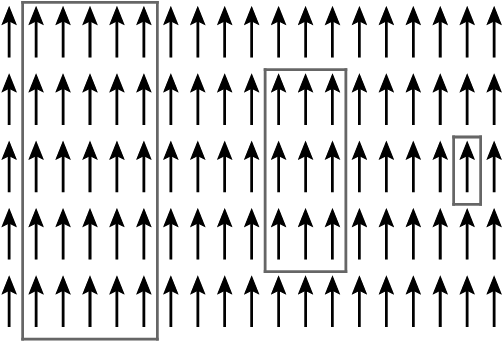}
\caption{Ordered state (self-similar).}\label{escala}
\end{figure}
\begin{figure}[H]
\centering
\includegraphics[width=0.45\textwidth]{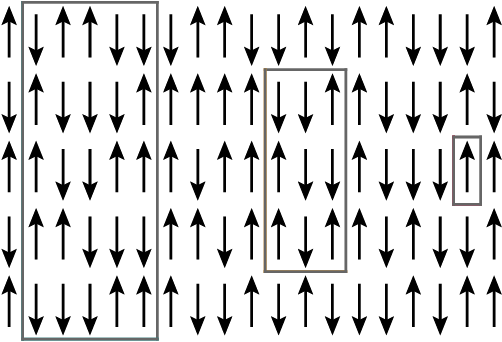}
\caption{Disordered state (not self-similar).}\label{escala2}
\end{figure}

\par Defining a theory capable of probing effects at all scales of length for a given system undergoing continuous phase transitions is mandatory \cite{WilsonSciAme}. In fact, for such a system, near the critical temperature, there is some ordered state and it is similar to itself or it is self-similar at all scales of length. In the case of magnetic systems, all spins are aligned as depicted in Figure \ref{escala}. While in the case of critical temperatures far away from the critical one, the system is is some disordered state and it is not self-similar at all scales of length. This situation is exemplified in Figure \ref{escala2}. 


\section{Nonideal statistical field theory}\label{Nonideal statistical field theory}

\par
\begin{figure}[H]
\centering
\includegraphics[width=0.43\textwidth]{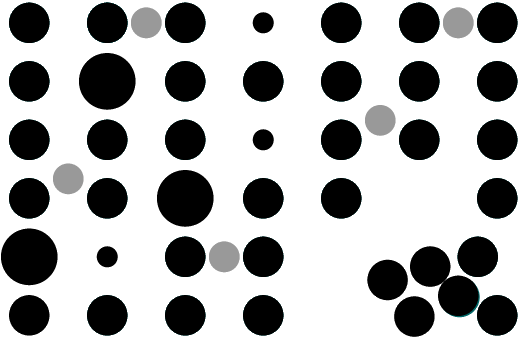}
\caption{Nonideal crystal.}\label{NonIdeal}
\end{figure}

\begin{figure}[h]
\centering
\includegraphics[width=0.43\textwidth]{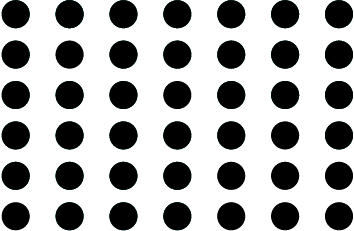}
\caption{Ideal crystal.}\label{Ideal}
\end{figure}

 We introduce the NISFT whose generating functional is given by
\begin{eqnarray}\label{huyhtrjisd}
Z[J] = \mathcal{N}^{-1}e_{a}^{-\int d^{d}x\mathcal{L}_{int}\left(\frac{\delta}{\delta J(x)}\right)}e^{\frac{1}{2}\int d^{d}xd^{d}x^{\prime}J(x)G_{0}(x-x^{\prime})J(x^{\prime})},
\end{eqnarray}
where $\mathcal{N}$ is a constant determined from the condition $Z[J=0] = 1$ and the nonideal distribution given by $a e_{a}^{x} = e^{ax} - 1 + a$ ($0 < a < 2$). The nonideal distribution reduces to the Boltzmann one in the limit $a\rightarrow 1$. The equation (\ref{huyhtrjisd}) describes the critical behavior whose situation ($a\neq 1$) is that of a nonideal system (can be a crystal) with defects (black dislocated atoms at the bottom right corner and vacancies), inhomogeneities (black bigger and smaller atoms) and impurities (grey atoms of another type) shown in Figure \ref{NonIdeal}. On the other hand, some ideal system ($a = 1$) regularly arranged with same atoms and of same size is represented in Figure \ref{Ideal}. The interactions in the ideal case, evidently, distribute according to the Boltzmann distribution for which $a\rightarrow 1$. In turn, the nonideal effects modify how the interactions distribute along the crystal and now they distribute in accordance with the nonideal distribution and we have to consider $a\neq 1$. So there is now some interplay between nonideal effects and ﬂuctuations and the aim of this work is to compute such fluctuations modifying the values of the nonideal critical exponents beyond LO or up to NLO.


\subsection{Nonideal critical exponents}

\par The nonideal critical exponents up to NLO computed through six distinct and independent methods are given by 
\begin{eqnarray}\label{etaa}
\eta_{a} =  \dfrac{(N + 2)}{2a(N + 8)^{2}}\epsilon^{2} + \dfrac{(N + 2)}{8a(N + 8)^{4}}(-N^{2} + 56N + 272)\epsilon^{3}, 
\end{eqnarray}
\begin{eqnarray}\label{nua}
\nu_{a} = \dfrac{1}{2} + \dfrac{(N + 2)}{4a(N + 8)}\epsilon + \dfrac{(N + 2)}{8a^{2}(N + 8)^{3}}[N^{2} + (19a + 4)N + 92a - 32]\epsilon^{2}.
\end{eqnarray}
Now we compare these values with experimental measured ones. 

\section{Comparison between theoretic and experimental results}\label{Comparison}

\subsection{Nonideal Ising systems}

\par The critical exponents for some nonideal materials were measured and are shown in Table \ref{tableIsing}. Their values deviate from those for ideal Ising systems, namely $\beta = 0.325(2)$ and $\gamma = 1.241(2)$ \cite{PHAN2010238}. The corresponding values, by employing the scaling relations $2\beta = \nu(d - 2 + \eta)$ and $\gamma = \nu(2 - \eta)$, from Eqs. (\ref{etaa})-(\ref{nua}) for $a = 1$, are $\beta = 0.325$ and $\gamma = 1.230$ \cite{Wilson197475}. The percentage relative errors to the six-loop values $\beta = 0.3260(5)$ and $\gamma = 1.2356(14)$ \cite{PhysRevD.96.036016} are $0.307~\%$ and $0.485~\%$, respectively. In literature, acceptable percentage relative errors are in the interval $< 5~\%$. Obtaining $\beta$ and $\gamma$ through scaling relations improves their precision if we take $\nu$ at $\mathcal{O}(\epsilon^{3})$ and $\eta$ at $\mathcal{O}(\epsilon^{4})$ instead both at $\mathcal{O}(\epsilon^{3})$ (see Appendix \ref{Remaining critical exponents} for the exponents $\beta$, $\gamma$ and $\alpha$ at $\mathcal{O}(\epsilon^{3})$ and $\delta$ at $\mathcal{O}(\epsilon^{4})$ for $a \neq 1$). For $a \neq 1$, we can apply the nonideal scaling relations $2\beta_{a} = \nu_{a}(d - 2 + \eta_{a})$ and $\gamma_{a} = \nu(2 - \eta_{a})$, which have the same functional form as those for ideal systems (it was shown in Ref. \cite{CARVALHO2023137683} that the corresponding scaling relations do not depend on the parameter of that Ref. and we can apply the same reasoning, without loss of generality, for the theory of this work), the values of Table \ref{tableIsing} range from 
\begin{eqnarray}
0.301~(a = 1.5) < \beta_{a} < 0.333~(a = 0.9)~\text{up to NLO},
\end{eqnarray}
\begin{eqnarray}
1.160~(a = 1.5) < \gamma_{a} < 1.251~(a = 0.9)~\text{up to NLO}.
\end{eqnarray}
By considering only LO terms, we obtain 
\begin{eqnarray}
0.297~(a = 1.001) < \beta_{a} < 0.335~(a = 0.56)~\text{up to LO},
\end{eqnarray}
\begin{eqnarray}
1.156~(a = 1.001) < \gamma_{a} < 1.276~(a = 0.56)~\text{up to LO}.
\end{eqnarray}

\begin{table}[H]
\caption{Critical exponents to $3$d ($\epsilon = 1$) Ising ($N = 1$) systems, obtained from experiment through Modified Arrott (MA) plots \cite{PhysRevLett.19.786}, Kouvel-Fisher (KF) method \cite{PhysRev.136.A1626}.}
\begin{tabular}{ p{5.4cm}p{1.1cm}p{1.1cm}  }
 \hline
 Ising & $\beta$ & $\gamma$    \\
 \hline
 Nd$_{0.55}$Sr$_{0.45}$Mn$_{0.98}$Ga$_{0.02}$O$_{3}$\cite{YU2018393}  & 0.308(10) &  1.197 \\
 Pr$_{0.6}$Sr$_{0.4}$MnO$_{3}$\cite{PhysRevB.92.024409}MAP  &   0.315(0) &  1.095(7)      \\
 Pr$_{0.6}$Sr$_{0.4}$MnO$_{3}$\cite{PhysRevB.92.024409}KF  &   0.312(2) &  1.106(5)      \\
 La$_{0.8}$Ca$_{0.2}$MnO$_{3}$\cite{ZHANG2013146}KF  &   0.316(7) &  1.081(36)      \\
 La$_{0.7}$Ca$_{0.3}$Mn$_{0.85}$Ni$_{0.15}$O$_{3}$\cite{PHAN201440}MAP  &   0.320(9) &  0.990(82)      \\
 Nd$_{0.6}$Sr$_{0.4}$MnO$_{3}$\cite{RSCAdvJeddi}MAP  &   0.320(6) &  1.239(2)      \\
 Nd$_{0.6}$Sr$_{0.4}$MnO$_{3}$\cite{RSCAdvJeddi}KF  &   0.323(2) &  1.235(4)      \\
 Nd$_{0.6}$Sr$_{0.4}$MnO$_{3}$\cite{PhysRevB.92.024409}MAP  &   0.321(3) &  1.183(17)      \\
 Nd$_{0.6}$Sr$_{0.4}$MnO$_{3}$\cite{PhysRevB.92.024409}KF  &   0.308(4) &  1.172(11)      \\
 Nd$_{0.67}$Ba$_{0.33}$MnO$_{3}$\cite{HCINI20152042}MAP  &   0.325(4) &  1.248(19)      \\
 Nd$_{0.67}$Ba$_{0.33}$MnO$_{3}$\cite{HCINI20152042}KF  &   0.326(5) &  1.244(33)      \\
 La$_{0.65}$Bi$_{0.05}$Sr$_{0.3}$MnO$_{3}$\cite{Phys.SolidStateBaazaoui}MAP  &   0.335(3) &  1.207(20)      \\
 La$_{0.65}$Bi$_{0.05}$Sr$_{0.3}$MnO$_{3}$\cite{Phys.SolidStateBaazaoui}KF  &   0.316(7) &  1.164(20)      \\
 La$_{0.65}$Bi$_{0.05}$Sr$_{0.3}$Mn$_{0.94}$Ga$_{0.06}$O$_{3}$\cite{Phys.SolidStateBaazaoui}MAP & 0.334(4) & 1.192(8) \\
 La$_{0.65}$Bi$_{0.05}$Sr$_{0.3}$Mn$_{0.94}$Ga$_{0.06}$O$_{3}$\cite{Phys.SolidStateBaazaoui}KF  & 0.307(8) & 1.138(5) \\
 \hline
 \end{tabular}
\label{tableIsing}
\end{table}

\subsection{Nonideal Heisenberg systems}

\par The same reasoning employed for Ising systems can be applied for Heisenberg ones of Table \ref{tableHeisenberg}. These values deviate from those for ideal Heisenberg measured systems, namely $\beta = 0.365(3)$ and $\gamma = 1.336(4)$ \cite{PHAN2010238}. The values of $\beta$ and $\gamma$ from Eqs. (\ref{etaa})-(\ref{nua}), from scaling relations, for $a = 1$ are $\beta = 0.352$ and $\gamma = 1.330$ \cite{Wilson197475}, respectively, whose percentage relative errors to the corresponding most recent six-loop values $\beta = 0.3663(12)$ and $\gamma = 1.385(4)$ \cite{PhysRevD.96.036016} are $3.825~\%$ and $4.068~\%$, respectively. Now for $a \neq 1$, we obtain that the values of the Table \ref{tableHeisenberg} range from 
\begin{eqnarray}
0.343~(a = 1.1) < \beta_{a} < 0.441~(a = 0.54)~\text{up to NLO},
\end{eqnarray}
\begin{eqnarray}
1.302~(a = 1.1) < \gamma_{a} < 1.585~(a = 0.54)~\text{up to NLO}.
\end{eqnarray}
If we compute the same exponents only up to LO we obtain 
\begin{eqnarray}
0.319~(a = 1.001) < \beta_{a} < 0.441~(a = 0.38)~\text{up to LO}),
\end{eqnarray}
\begin{eqnarray} 
1.214~(a = 0.001) < \gamma_{a} < 1.555~(a = 0.38)~\text{up to LO}.
\end{eqnarray}
We see that, for both Ising and Heisenberg nonideal systems, different number of loops considered lead to distinct values of $a$. So when we take more and more loops, we get closer and closer to the true value of $a$.

\begin{table}[H]
\caption{Critical exponents to $3$d ($\epsilon = 1$) Heisenberg ($N = 3$) systems, obtained from experiment through Modified Arrott (MA) plots \cite{PhysRevLett.19.786}, Kouvel-Fisher (KF) method \cite{PhysRev.136.A1626}.}
\begin{tabular}{ p{5.3cm}p{1.1cm}p{1.1cm}  }
 \hline
 Heisenberg & $\beta$ & $\gamma$    \\
 \hline
 Pr$_{0.77}$Pb$_{0.23}$MnO$_{3}$\cite{PhysRevB.75.024419}MAP  &   0.343(5) &  1.357(20)     \\
 Pr$_{0.77}$Pb$_{0.23}$MnO$_{3}$\cite{PhysRevB.75.024419}KF  &   0.344(1) &  1.352(6)     \\
 AMnO$_{3}$\cite{OMRI20123122}MAP  &   0.355(7) &  1.326(2)     \\
 AMnO$_{3}$\cite{OMRI20123122}KF  &   0.344(5) &  1.335(2)     \\
 Nd$_{0.7}$Pb$_{0.3}$MnO$_{3}$\cite{Ghosh_2005}MAP  &   0.361(13) &  1.325(1)           \\
 Nd$_{0.7}$Pb$_{0.3}$MnO$_{3}$\cite{Ghosh_2005}KF  &   0.361(5) &  1.314(1)           \\
 LaTi$_{0.2}$Mn$_{0.8}$O$_{3}$\cite{doi:10.1063/1.2795796}KF  &   0.359(4) &  1.280(10)     \\
 La$_{0.67}$Sr$_{0.33}$Mn$_{0.95}$V$_{0.05}$O$_{3}$\cite{MNEFGUI2014193}MAP  &   0.358(5) &  1.381(4)  \\
 Nd$_{0.85}$Pb$_{0.15}$MnO$_{3}$\cite{Ghosh_2005}MAP  &  0.372(1) &  1.340(30)  \\
 Nd$_{0.85}$Pb$_{0.15}$MnO$_{3}$\cite{Ghosh_2005}KF  &  0.372(4) &  1.347(1)  \\
 Nd$_{0.6}$Pb$_{0.4}$MnO$_{3}$\cite{PhysRevB.68.144408}KF  &   0.374(6) &  1.329(3)  \\
 La$_{0.67}$Sr$_{0.33}$Mn$_{0.95}$V$_{0.15}$O$_{3}$\cite{MNEFGUI2014193}MAP  &   0.375(3) &  1.355(6)  \\
 La$_{0.67}$Ba$_{0.22}$Sr$_{0.11}$MnO$_{3}$\cite{BenJemaa}MAP  &   0.378(3) &  1.388(1)   \\
 La$_{0.67}$Ba$_{0.22}$Sr$_{0.11}$MnO$_{3}$\cite{BenJemaa}KF  &   0.386(6) &  1.393(4)   \\
 LaTi$_{0.95}$Mn$_{0.05}$O$_{3}$\cite{doi:10.1063/1.2795796}KF  &   0.378(7) &  1.290(20)     \\
 LaTi$_{0.9}$Mn$_{0.1}$O$_{3}$\cite{doi:10.1063/1.2795796}KF  &   0.375(5) &  1.250(20)     \\
 LaTi$_{0.85}$Mn$_{0.15}$O$_{3}$\cite{doi:10.1063/1.2795796}KF  &   0.376(3) &  1.240(10)     \\
 La$_{0.67}$Ca$_{0.33}$Mn$_{0.9}$Ga$_{0.1}$O$_{3}$\cite{PhysRevB.70.104417}MAP  &  0.380(2) &  1.365(8)  \\
 La$_{0.67}$Ca$_{0.33}$Mn$_{0.9}$Ga$_{0.1}$O$_{3}$\cite{PhysRevB.70.104417}KF  &   0.387(6) &  1.362(2)  \\
 La$_{0.67}$Ba$_{0.22}$Sr$_{0.11}$Mn$_{0.9}$Fe$_{0.1}$O$_{3}$\cite{BenJemaa}MAP  &   0.398(2) &  1.251(5)   \\
 La$_{0.67}$Ba$_{0.22}$Sr$_{0.11}$Mn$_{0.9}$Fe$_{0.1}$O$_{3}$\cite{BenJemaa}KF  &   0.395(3) &  1.247(3)   \\
 La$_{0.67}$Ba$_{0.22}$Sr$_{0.11}$Mn$_{0.8}$Fe$_{0.2}$O$_{3}$\cite{BenJemaa}MAP  &   0.411(1) &  1.241(4)   \\
 La$_{0.67}$Ba$_{0.22}$Sr$_{0.11}$Mn$_{0.8}$Fe$_{0.2}$O$_{3}$\cite{BenJemaa}KF  &   0.394(3) &  1.292(3)   \\
 Pr$_{0.7}$Pb$_{0.3}$MnO$_{3}$\cite{PhysRevB.75.024419}MAP  &   0.404(6) &  1.354(20)     \\
 Pr$_{0.7}$Pb$_{0.3}$MnO$_{3}$\cite{PhysRevB.75.024419}KF  &   0.404(1) &  1.357(6)     \\
 La$_{0.75}$(Sr,Ca)$_{0.25}$Mn$_{0.9}$Ga$_{0.1}$O$_{3}$\cite{BenJemaa}MAP  &   0.420(5) &  1.221(2)   \\
 La$_{0.75}$(Sr,Ca)$_{0.25}$Mn$_{0.9}$Ga$_{0.1}$O$_{3}$\cite{BenJemaa}KF  &   0.428(5) &  1.286(4)   \\
 Pr$_{0.5}$Sr$_{0.5}$MnO$_{3}$\cite{PhysRevB.79.214426}MAP  &   0.443(2) &  1.339(6)     \\
 Pr$_{0.5}$Sr$_{0.5}$MnO$_{3}$\cite{PhysRevB.79.214426}KF  &   0.448(9) &  1.334(10)     \\
 \hline
\end{tabular}
\label{tableHeisenberg}
\end{table}

\section{Conclusions}\label{Conclusions}

\par In this work we have introduced a field theory capable of describing the critical properties of nonideal systems undergoing continuous phase transitions up to NLO by improving earlier results of effective field theories \cite{CARVALHO2024139487,CARVALHO2023138187,ALVES2023138005,CARVALHO2023137683} limited only up to LO. These systems present defects, inhomogeneities and impurities as opposed to ideal ones which are perfect, homogeneous and pure. The interactions in the ideal case distribute according to the Boltzmann distribution for which $a\rightarrow 1$. In turn, the nonideal effects modify how the interactions distribute along the crystal and now they distribute in accordance with the nonideal distribution and we have to consider $a\neq 1$. The physical interpretation of the results of the present work is as follows: as an example, the nonideal susceptibility scaling critical behavior is governed by the nonideal critical exponent $\gamma_{a}$, namely, $\chi_{a}\propto |T - T_{c}|^{-\gamma_{a}}$. The susceptibility displays stronger (weaker) divergence for $a < 1$ ($a > 1$) when the critical temperature approaches the critical one, once the values of $\gamma_{a}$ get higher (lower). Then higher (lower) values of the critical exponents correspond to more (less) susceptible or weakly (strongly) interacting physical systems. So there is now some interplay between nonideal effects and ﬂuctuations and the aim of this work was to compute such fluctuations modifying the values of the nonideal critical exponents up to NLO. We have computed the values of the critical exponents of the theory up to NLO and have compared with their corresponding experimental measured results in Tables \ref{tableIsing} - \ref{tableHeisenberg}. The results were satisfactory and within the margin of error. We have shown that different number of loops considered has led to distinct values of $a$. So when we take more and more loops, we get closer and closer to the true value of $a$. Previous effective field theories taking into account nonideal critical properties just at LO were defined \cite{CARVALHO2024139487,CARVALHO2023138187,ALVES2023138005,CARVALHO2023137683}. As they did not describe nonideal results beyond LO, they must be discarded as fundamental ones, \emph{i. e.}, as ones representing defects, inhomogeneities and impurities taken at all scales of length and have to be employed at most as effective theories.

\appendix 

\section{Critical exponents for some effective theories}\label{Critical exponents for some effective theories}
\begin{table}[H]
\caption{Critical exponents for some effective theories.}
\begin{tabular}{p{8.0cm}p{8.0cm}}
\hline
 $\eta$ & $\nu$    \\
\hline
\hspace{.1mm} \\
\cite{CARVALHO2024139487} $\eta_{\delta_{KLS}} = \dfrac{1}{1 - 2\delta_{KLS}}\dfrac{(N + 2)}{2(N + 8)^{2}}\epsilon^{2}$ &  \cite{CARVALHO2024139487} $\nu_{\delta_{KLS}} = \dfrac{1}{2} + \dfrac{1}{1 - 2\delta_{KLS}}\dfrac{(N + 2)}{4(N + 8)}\epsilon$ \\
\hspace{.1mm} \\
\cite{CARVALHO2023138187} $\eta_{\gamma_{KLS}} = \dfrac{1}{1 - \gamma_{KLS}}\dfrac{(N + 2)}{2(N + 8)^{2}}\epsilon^{2}$ & \cite{CARVALHO2023138187} $\nu_{\gamma_{KLS}} = \dfrac{1}{2} + \dfrac{1}{1 - \gamma_{KLS}}\dfrac{(N + 2)}{4(N + 8)}\epsilon$   \\
\hspace{.1mm} \\
\cite{ALVES2023138005} $\eta_{\kappa} = \dfrac{(N + 2)}{2(N + 8)^{2}}\epsilon^{2}$ & \cite{ALVES2023138005} $\nu_{\kappa} = \dfrac{1}{2} + \dfrac{(N + 2)}{4(N + 8)}\epsilon$ \\
\hspace{.1mm} \\
\cite{CARVALHO2023137683} $\eta_{q} = \dfrac{1}{q}\dfrac{(N + 2)}{2(N + 8)^{2}}\epsilon^{2}$ & \cite{CARVALHO2023137683} $\nu_{q} = \dfrac{1}{2} + \dfrac{1}{q}\dfrac{(N + 2)}{4(N + 8)}\epsilon$ \\
\hspace{.1mm} \\
\cite{Shafee} $\eta_{q} = \dfrac{1}{3 - 2q}\dfrac{(N + 2)}{2(N + 8)^{2}}\epsilon^{2}$ & \cite{Shafee} $\nu_{q} = \dfrac{1}{2} + \dfrac{1}{3 - 2q}\dfrac{(N + 2)}{4(N + 8)}\epsilon$ \\
\hspace{.1mm} \\
\cite{Fibonacci} $\eta_{q_{1},q_{2}} = \dfrac{2}{q_{1}^{2} + q_{2}^{2}}\dfrac{(N + 2)}{2(N + 8)^{2}}\epsilon^{2}$ & \cite{Fibonacci} $\nu_{q_{1},q_{2}} = \dfrac{1}{2} + \dfrac{2}{q_{1}^{2} + q_{2}^{2}}\dfrac{(N + 2)}{4(N + 8)}\epsilon$ \\
\hspace{.1mm} \\
\cite{Jackson} $\eta_{q} = \dfrac{2}{1 + q}\dfrac{(N + 2)}{2(N + 8)^{2}}\epsilon^{2}$ & \cite{Jackson} $\nu_{q} = \dfrac{1}{2} + \dfrac{2}{1 + q}\dfrac{(N + 2)}{4(N + 8)}\epsilon$ \\
\hspace{.1mm} \\
\hline
\end{tabular}
\label{CENRT}
\end{table}

\section{Remaining critical exponents}\label{Remaining critical exponents}
\begin{eqnarray}\label{alphaa}
\alpha_{a} =  - \dfrac{[(2 - a)N - 4(2a - 1)]}{2a(N + 8)^{2}}\epsilon  - \dfrac{(N + 2)}{4a^{2}(N + 8)^{3}}[(2 - a)N^{2} + (22a + 8)N - 120a + 64]\epsilon^{2},
\end{eqnarray}
\begin{eqnarray}\label{betaa}
\beta_{a} =  \dfrac{1}{2} - \dfrac{[4a - 1 + (a - 1)N]}{2a(N + 8)}\epsilon + \dfrac{(N + 2)}{2a^{2}(N + 8)^{3}}\left[\dfrac{(1 - a)N^{2}}{4} + (a + 1)N + 9a - 8\right]\epsilon^{2},
\end{eqnarray}
\begin{eqnarray}\label{gammaa}
\gamma_{a} = 1 + \dfrac{(N + 2)}{2a(N + 8)}\epsilon + \dfrac{(N + 2)}{4a^{2}(N + 8)^{3}}[N^{2} + (18a + 4)N + 84a - 32]\epsilon^{2},
\end{eqnarray}
\begin{eqnarray}\label{deltaa}
&& \delta_{a} = 3 + \epsilon + \dfrac{1}{2a(N + 8)^{2}}[aN^{2} + (16a - 2)N + 64a - 4]\epsilon^{2} + \nonumber \\ && \dfrac{1}{4a(N + 8)^{4}}[aN^{4} + (29a + 1)N^{3} (330a - 54)N^{2} + (1760a - 384)N + 3712a - 544]\epsilon^{3}.
\end{eqnarray}

\textbf{Acknowledgements} PRSC would like to thank the Brazilian funding agencies CAPES and CNPq (Grant: Produtividade 306130/2022-0) for financial support.

\bibliography{apstemplate}

\end{document}